\documentclass[aps,prl,twocolumn,superscriptaddress,groupedaddress,a4paper]{revtex4}
\usepackage{graphicx}
\usepackage{color}
\usepackage{appendix}
\begin{document}

\title{Self-organized defect strings in two-dimensional crystals}

\author{Wolfgang Lechner}
\affiliation{Institute for Quantum Optics and Quantum Information and Institute for Theoretical Physics, University of Innsbruck, 6020 Innsbruck, Austria}

\author{David Polster}
\affiliation{Department of Physics, University of Konstanz, D-78457 Konstanz, Germany}

\author{Georg Maret}
\affiliation{Department of Physics, University of Konstanz, D-78457 Konstanz, Germany}

\author{Peter Keim}
\affiliation{Department of Physics, University of Konstanz, D-78457 Konstanz, Germany}

\author{Christoph Dellago}
\affiliation{Faculty of Physics, University of Vienna, Boltzmanngasse 5, 1090 Vienna, Austria}

\date{\today}

\begin{abstract}
Using experiments with single particle resolution and computer simulations we study the collective behaviour of multiple vacancies injected into two-dimensional crystals. We find that the defects assemble into linear strings that propagate through the crystal in a succession of rapid one-dimensional gliding phases and rare rotations, during which the direction of motion changes. At both ends, strings are terminated by dislocations with anti-parallel Burgers vectors. By monitoring the separation of the dislocations, we measure their effective interactions with high precision, for the first time beyond spontaneous formation and annihilation, and explain the double-well form of the dislocation interaction in terms of continuum elasticity theory. Our results give a detailed picture of the motion and interaction of dislocations in two dimensions and enhance our understanding of topological defects in two-dimensional nano-materials. 
\end{abstract}

\maketitle

The plasticity and mechanical strength of solids is essentially governed by the dynamics of dislocations, topological defects generated during plastic deformation \cite{TAYLOR}. Recent research has focused on the properties and effects of such defects in novel two-dimensional nano-materials. In graphene, for instance, dislocations can significantly alter the electronic and mechanical properties of the material, and the ability to control them will be key for technological application  \cite{DEFECTS_GRAPHENE,DISLOCATION_GRAPHENE,LOUIE2010}. Dislocations also play a central role in two-dimensional melting, which, according to the Kosterlitz-Thouless-Halperin-Nelson-Young theory \cite{KTHNY},  is mediated by the formation and subsequent dissociation of dislocation pairs leading to the loss of quasi-long range translational order \cite{MARET,MARET1999,MELTINGREVIEW,KRAUTH}. Due to the remarkable experimental advances of recent years, such as the development of optical tweezers and confocal microscopy \cite{CONFOCAL,PERTSINIDIS_PRL,PERTSINIDIS_NJP,SCIENCE_GASSER,MARET,SPAEPEN_2006,PC2012}, the direct observation of the structure and dynamics of defects  is now possible with single particle resolution. However, measuring the interaction of dislocations experimentally for a wide range of distances and angles is challenging, mainly due to the lack of a systematic way to generate and control dislocations beyond their spontaneous formation and annihilation \cite{MARET}. 

\begin{figure*}[tb]
\centerline{\includegraphics[height=9cm]{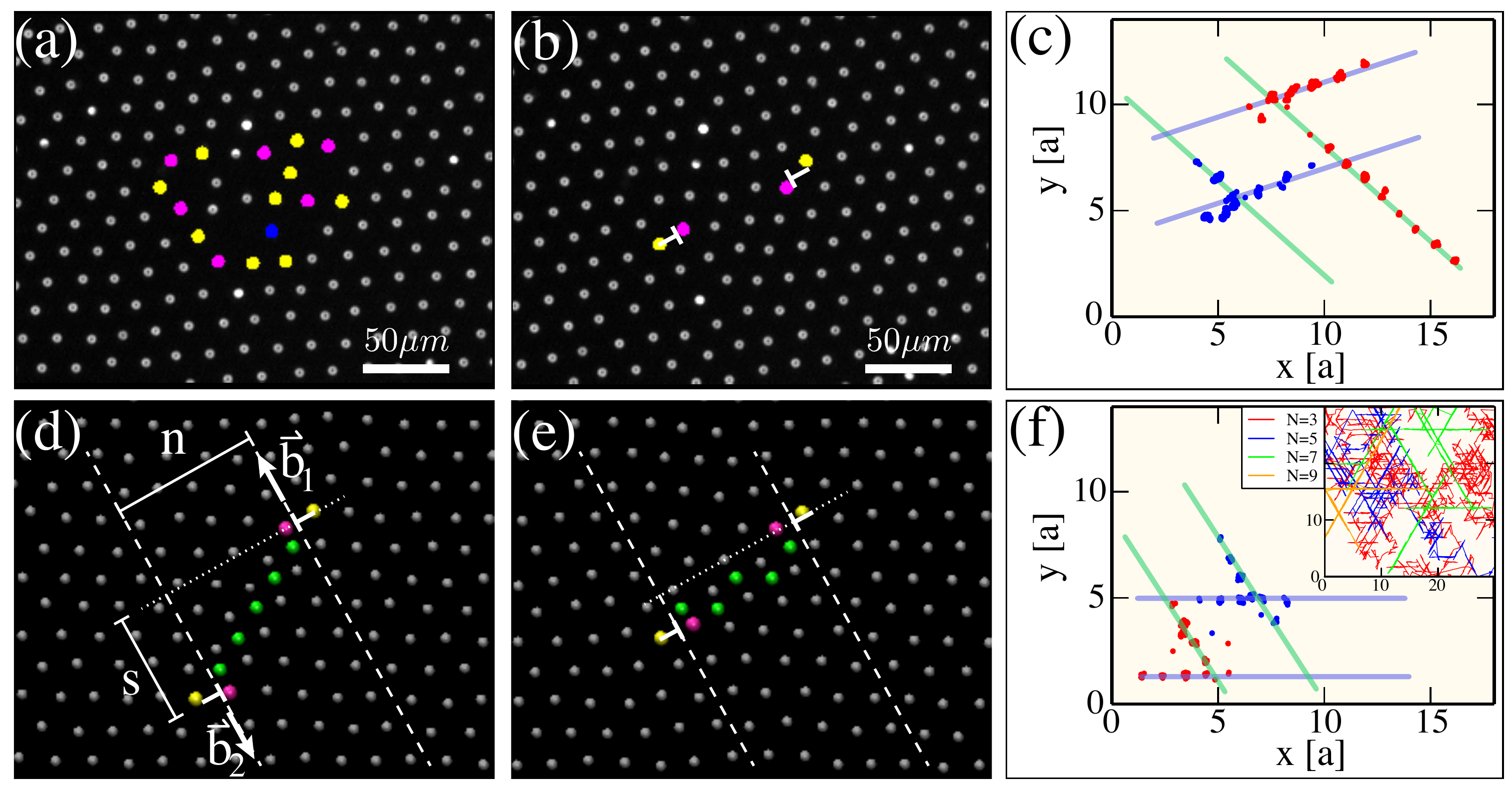}}
\caption{{\bf String formation and dislocation motion from experiment (top row) and simulation (bottom row)}. (a) Video-microscopy image of the colloidal crystal with $N=5$ vacancies. Colors distinguish particles with $4$ neighbors (blue), $5$ neighbors (yellow), $6$ neighbors (gray) and 7 neighbors (pink).  (b) After about 6 minutes the vacancies have coalesced into a vacancy string terminated by two dislocations. Each dislocation corresponds to a 5-7 coordinated particle pair and is indicated by a white T-symbol. Panels (d) and (e) are snapshots taken from Monte Carlo simulations performed at the conditions of the experiments. The two dislocations diffuse along parallel lattice lines indicated by dashed lines. The Burgers vectors ${\bf b}_1$ and ${\bf b}_2$ of the two dislocations have opposite direction and are parallel to the direction of motion. While the perpendicular separation $n$ between the dislocations is fixed by the number $N$ of vacancies introduced into the system, the parallel distance $s$ changes as the dislocations move under the influence of their effective interaction. The green particles mark the positions of the vacancies. Panels (c) and (f) display the positions of the dislocations, shown as red and blue dots, taken from two trajectories recorded for string length $N=4$ in the experiment (c) and in the computer simulation (f), respectively. In both cases, the dislocations first move along two parallel lines in one of the main lattice directions and then change their direction of motion during a rapid reorientation event that turns the Burgers vectors of the dislocations by $\pi/3$. The inset of panel (f) shows dislocation trajectories obtained from the computer simulations for various vacancy numbers $N$.
}
\label{fig:string}
\end{figure*}

In this Letter, we exploit the properties of self-organized defect strings to measure the interaction of dislocations using a planar crystal of colloidal particles with tuneable interactions as model system. As shown in Fig. \ref{fig:string}, multiple vacancies introduced artificially into the system rapidly cluster, driven by the attractive defect interaction \cite{TROYER, LDSoftMatter1, LDSoftMatter2}. As predicted by previous computer simulations \cite{LDSoftMatter2}, the vacancy clusters arrange into linear strings terminated by two dislocations of anti-parallel Burgers vectors. These defect strings are the two-dimensional analogs of the prismatic dislocation loops \cite{HIRTH_LOTHE,ZINKLE,ARAKAWA,DISLOCATIONS} forming in irradiated three-dimensional materials as a result of defect clustering. In contrast to dislocation pairs forming spontaneously by thermal excitation, the dislocations resulting from defect clustering do not annihilate due the geometrical constraints imposed by the defect string. While the separation of the dislocations perpendicular to the Burgers vectors is fixed and depends on the number of vacancies in the string, the dislocations are mobile in the parallel direction. As discussed below, we use the statistics of this motion, collected for various string lengths, to reconstruct the complete interaction Hamiltonian of dislocation pairs with anti-parallel Burgers vectors. In particular, we find an interaction of double well form, as predicted by elasticity theory. 

Our system consists of a monolayer of superparamagnetic colloidal particles confined by gravity to a flat water-air interface \cite{MARET1997,KEIM_RSI_2009}. A magnetic field of strength $B$ perpendicular to the interface induces magnetic dipole moments of magnitude $\chi B$, leading to the dipolar pair interaction
\begin{equation}
\label{equ:pairinteraction}
\beta v(r) =  \frac{3^{3/4}\Gamma}{(2\pi)^{3/2}} \left(\frac{a}{r}\right)^3.
\end{equation}
Here, $\beta = 1/k_B T$ with the Boltzmann constant $k_B$ and temperature $T$, $r$ is the particle-particle distance, $a$ is the lattice spacing of the perfect crystal and $\Gamma = \beta (\mu_0/4\pi)(\chi B)^2 (\pi \rho)^{3/2}$,  where $\mu_0$ is the permeability of vacuum and $\rho$ is the area density of the crystal. The dimensionless parameter $\Gamma$, which acts as an inverse temperature and can be tuned by varying the magnetic field, completely determines the structural properties of the system. Vacancies are generated by a fiber-coupled optical tweezer, where the objective can be moved in $x$-, $y$- and $z$-direction for distances larger than the field of view. The colloids are trapped in the laser focus for moderate field strength, but for high field strength (500mW, 100er tweezer-objective, n.a. 0.73, $\textrm{Ar}^{+}$-Laser) light pressure is strong enough to push particles through the interface, removing them from the monolayer. We also study the vacancy strings using $NVT$ Monte Carlo (MC) simulations with local trial moves (Methods).

In Figs. \ref{fig:string}(a) and (b) (experiment) and Figs. \ref{fig:string}(d) and (e) (simulation) we show dislocations and vacancies for typical defect strings. Dislocations are identified from a Voronoi construction and vacancies from a comparison with  an underlying undistorted lattice (see Supplementary Information). Even though there are $5$ vacancies present in the system, only two dislocations remain in the equilibrated string configuration. The region between the two dislocations is a prefect crystal, in which each particle has exactly $6$ neighbors. Nevertheless, through comparison with the underlying ideal lattice one can locate the entire string of vacancies that connect the dislocations at the string endpoints. 

\begin{figure}[tb]
\centerline{\includegraphics[width=7.5cm]{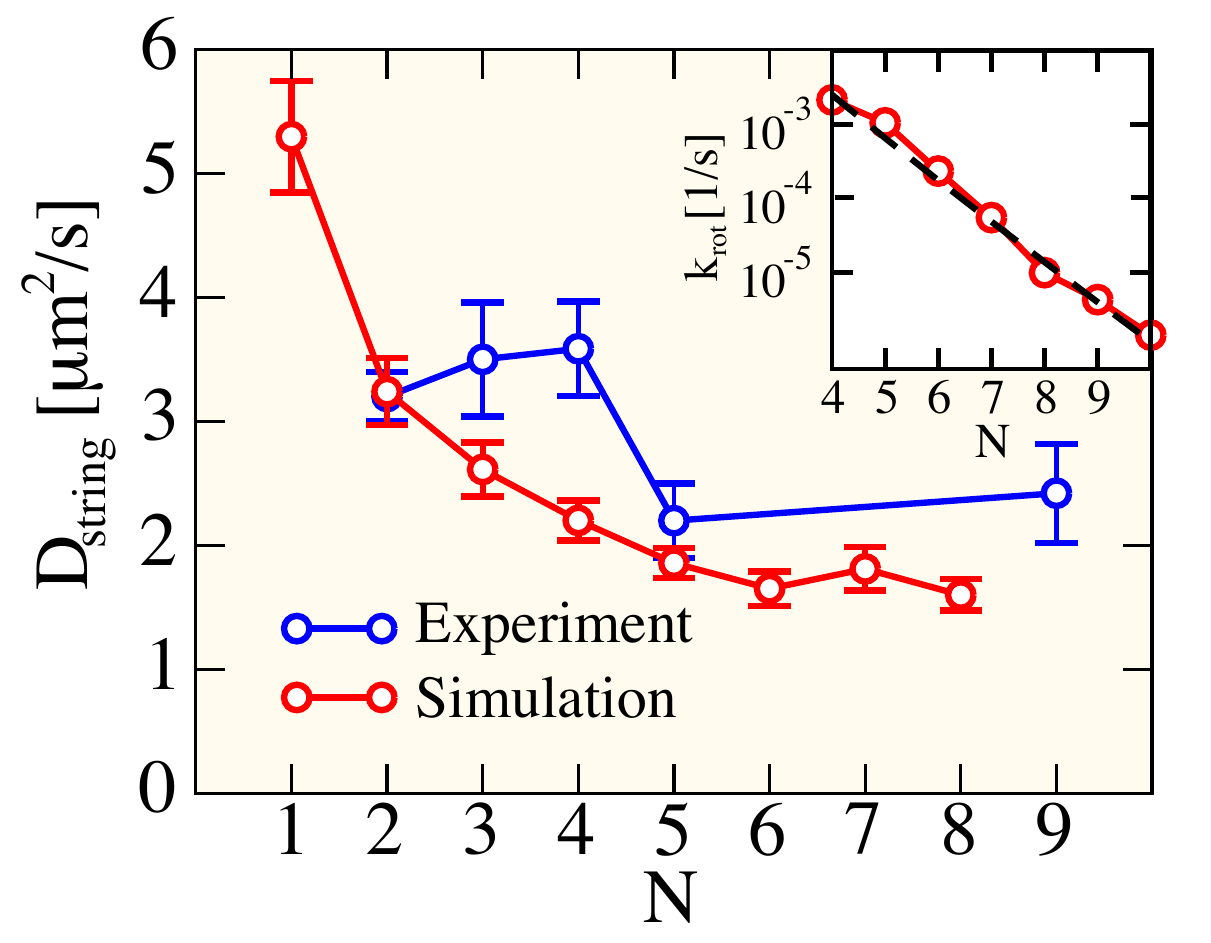}}
\caption{{\bf Diffusion constant $D_{\rm string}$ of vacancy strings as a function of defect number $N$}. Experimental results (blue) are shown together with results of the simulations (red). In the simulations, averages and error bars are calculated from $12$ blocks of $4 \times 10^6$ MC sweeps each. The inset shows the rotation rate $k_{\rm rot}$ (see Supplementary Information) of vacancy strings as function of $N$. A fit of $k(N) \propto \exp(-\beta N e_a)/N$, shown as dashed line, yields an activation energy per defect of  $e_a=1.13 k_{\rm B} T$.
}
\label{fig:diffrot}
\end{figure}

Analyzing trajectories obtained from our video microscopy experiments and computer simulations in terms of defect positions, we find that defect strings propagate through the system in a sequence of fast gliding motions and rare rotations (see Fig. \ref{fig:string} and videos in the Supplementary Information). During the gliding phase, the vacancies and dislocations at the string endpoints move in a direction that is parallel to the Burgers vectors of the dislocations and coincides with one of the three symmetry axes of the crystal. In the direction perpendicular to the direction of motion, distances between defects remain constant. As a consequence, the string endpoints move on parallel lines, separated by $N$ lattice rows. In this gliding phase, the motion of the defect string can be viewed as the diffusion of the two endpoint dislocations along parallel lines under the influence of their mutual effective interaction. In the direction of motion the dislocations interact attractively at large distances, such that the two dislocations can diffuse only as a coupled pair. On short time scales, the dislocations thus diffuse freely with diffusion constant $D_{\text{\rm dis}}$, while at long time scales the diffusion of the dislocations is governed by the diffusion constant $D_{\text{\rm string}} = D_{\text{dis}}/2$ \cite{CHRISTOPHPROTONS}. 
Diffusion constants obtained from computer simulations and experiments, depicted in Fig. \ref{fig:diffrot}, agree within the statistical accuracy of the results. Remarkably, the diffusion constant is independent of the string length for $N > 5$, corroborating the view of the string motion as the diffusion of two coupled dislocations. Thus, the number of vacancies in the string influences the time at which the long-time diffusion regime is reached, but not the diffusion constant itself \cite{CHRISTOPHPROTONS}. 

\begin{figure}[tb]
\centerline{\includegraphics[width=8cm]{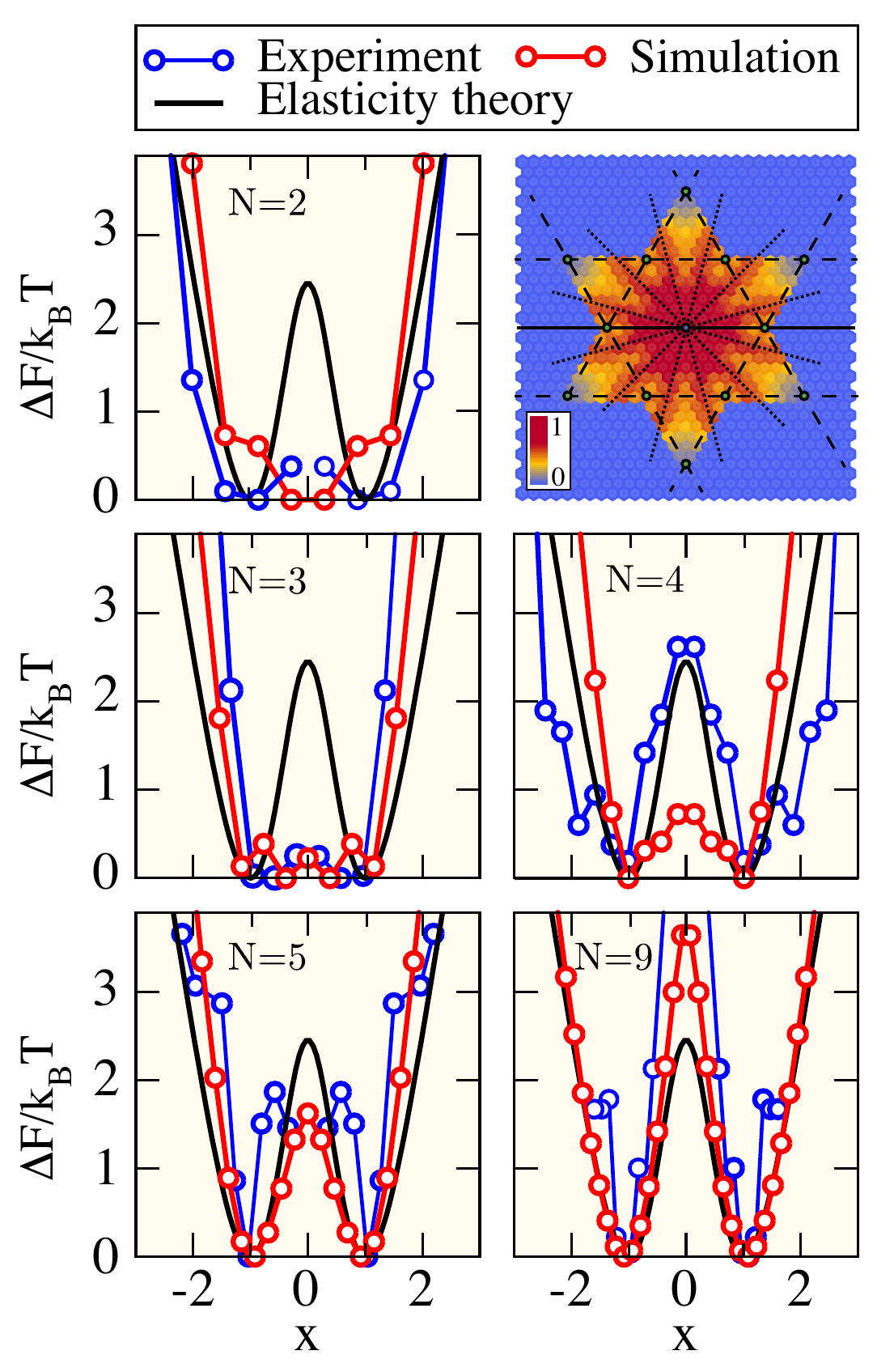}}
\caption{{\bf Defects interactions.} Free energy $\beta \Delta F$ as a function of the scaled distance $x=2s/(\sqrt{3}N)$ in the direction of motion for defect strings with lengths  from $N=2$, 3, 4, 5 and 9. Experimental results are shown together with results of simulations and the prediction of elasticity theory, Equ. (\ref{equ:endtoendscaled}). The top right inset depicts a color coded histogram of vacancy locations with for $N=8$ and one string-end at the origin (blue dot). The vacancy string follows the same double well pattern with angle $\pi/4$ (dotted lines) as the endpoints. The relative distance between dislocations (dashed lines) is constrained to $6$ lines with $12$ rotation points (green dots) where the string can change the gliding orientation. 
}
\label{fig:doublewell}
\end{figure}

The phases of facile gliding motion are interrupted by rotation events that change the direction of motion of the dislocations allowing the diffusion to become two-dimensional. Rotations can take place only at specific rotation points at the intersections of the lattice lines indicated by dashed lines in Fig. \ref{fig:doublewell} (top right), along which the endpoint dislocations have constant perpendicular distance. During such a rotation event, which can be detected based on the directions of the Burgers vectors or the sequence of dislocation positions (Methods), the string changes from one of the $3$ possible directions of motion to another. Examples of two rotation events observed in the experiments and computer simulations are shown in Fig. \ref{fig:string}, panels (c) and (f), respectively. Since rotation events are rare, we could determine the rotation rate reliably only for the long trajectories obtained from our computer simulations. As shown in the inset of Fig. \ref{fig:diffrot}, the rotation rate $k_{\rm rot}$, defined as the number of rotation events per second, decreases rapidly with string length. The particular functional form of $k_{\rm rot}(N)$ can be understood as follows. The probability to observe a rotation at a given time is the product of the probability to be at a rotation point times the probability to change the direction of motion at this point. The probability to be at a rotation point decreases as $1/N$ simply because of the number of locations available to the string endpoints is proportional to $N$. At a rotation point, a string can change its direction by a collective slip of two rows of $N$ particles past each other in the region between the endpoint dislocations. This collective rearrangement involves an activation energy that increases linearly with the string length, $E_a = N e_a$. Including both contributions to the rotation probability, one expects  a rotation rate $k_{\rm rot}(N) \propto \exp(-\beta N e_a)/N$. A fit of this expression to the simulation results yields good agreement, as shown in the inset of Fig. \ref{fig:diffrot}. 

To explore the effective interaction between the two dislocations located at the endpoints of the strings, we have determined the probability $P(s)$ of finding the dislocations at distance $s$ in the direction of motion [see Fig. \ref{fig:string}(d)]. The probabilities $P(s)$ are obtained by histogramming the distances $s$ along trajectories recorded in the experiments and in the simulations. We ignore configurations in which additional dislocation pairs arise through thermal fluctuations ($10$-$20\%$ of the configurations) and assume that the spontaneous formation of additional dislocations does not influence the defect string configuration. Therefore, neglecting these multi-dislocation configurations does not affect the calculated effective interactions given by $F(s)= -k_{\rm B}T\ln P(s)$. In Fig. \ref{fig:doublewell}, effective interactions $\beta F$, displayed as a function of the scaled coordinate $x = 2 s /(\sqrt{3} N)$, are compared to the predictions of elasticity theory for various string lengths $N$. The effective interactions obtained from our experiments agree well with the results of the computer simulations with deviations that are most likely due to the limited statistics of the experimental results. 

For strings longer than $N > 3$, both experiments and simulations yield effective interactions of double-well form with minima at $x=\pm 1$ corresponding to $s=\pm \sqrt{3} N/2$, and a central barrier at $x=0$. As a consequence, for all string lengths the defect strings are preferentially oriented at an angle of $\pi/4$ with respect to the high symmetry directions, and also individual defects follow this $\pi/4$-pattern (see Fig. \ref{fig:doublewell} top right). The double-well form of the dislocation interaction and, in particular, the surprising defect string alignment, differing from the high symmetry directions of the hexagonal lattice, can be rationalized in terms of continuum elasticity theory. In two dimensions, two dislocations with Burgers vectors ${\bf b}_1$ and ${\bf b}_2$ and separated by ${\bf R}$ are predicted to interact with energy $\beta F = -(K/4 \pi) [({\bf b}_1 \cdot {\bf b}_2) \ln R - ({\bf b}_1 \cdot {\bf R})({\bf b}_2 \cdot {\bf R}) /R^2]$, where $K$ is Young's modulus \cite{NABARRO,YOUNGMODULUS}. For a string of $N$ vacancies, the geometry of the dislocations can be described by ${\bf b}_1 = (-1,0)$, ${\bf b}_2 = (1,0)$ and ${\bf R} = (s,N \sqrt{3}/2)$, yielding the effective interaction 
\begin{equation}
\label{equ:endtoendscaled}
\beta F(x) = \frac{K}{8 \pi} \left(\frac{1-x^2}{1+x^2} +  \ln \frac{1+x^2}{2} \right).
\end{equation}
Note that this expression is valid for an isotropic system irrespective of the symmetry of the underlying crystal lattice to which no reference is made. The  free energy of Equ. (\ref{equ:endtoendscaled}) is a symmetric double well with minima at $x= \pm 1$ separated by a barrier of height $h = \beta F(0)  =  (K/8\pi) (1-\ln 2)$ located at $x=0$.  Thus, the barrier height depends on Young's modulus $K$ but not on the string length $N$. As can be seen in Fig. \ref{fig:doublewell}, for sufficiently long defect strings the free energy obtained from elasticity theory reproduces both the experimental and numerical results  to a remarkable extent. In particular, the positions of the minima, which are related to the preferential $\pi/4$-orientation of the strings, are predicted very accurately by elasticity theory for $N>3$. For short strings, elasticity theory fails at short distances since the discrete lattice becomes noticeable and for $N=2$ the potential becomes a single well. While the general shape of the free energy and the position of the minima are predicted well by elasticity theory, the predicted height of the barrier at $x=0$ differs significantly from the simulation results. Barrier heights obtained in our simulations converge to a constant value for growing string length as expected from elasticity theory. This height, however, exceeds the prediction of linear elasticity by almost 50\%, possibly due to non-linear elastic interactions. 

In summary, we have used a combination of experiment and computer simulations to characterize the motion and interactions of defects in two-dimensional crystals with single particle resolution. We have found that vacancies introduced artificially into the system coalesce into string-like clusters that are almost as mobile as single vacancies. The dynamics of these strings consists of long stretches of fast diffusive gliding in one dimension punctuated by rotations at specific rotation points. For rotations a collective rearrangement, reorienting the dislocations terminating the vacancy string is needed. From the statistics of dislocation positions tracked with video microscopy we reconstruct the effective interactions between dislocations in two dimensions and the double well predicted by continuum elasticity. Based on our computer simulations, we predict that analogous clustering phenomena occur for interstitial introduced into 2d crystals, providing another way to study defect clustering and dislocation interactions in 2d materials. In future work we will study defect clustering and interactions in binary systems as well as in systems with anisotropic interactions which can be realised in our model system by tilting the magnetic field controlling the interactions between particles. Approaches analogous to the ones used here to explore defects in colloidal crystals could be applied to directly probe dislocation interactions in two-dimensional nano-materials such as graphene or boron nitride sheets in high-resolution transmission electron microscopy experiments \cite{DISLOCATION_GRAPHENE}. 

This research was supported by the Austrian Science Fund (FWF) within the SFB ViCoM (Projekt  F41) and P 25454-N27. Simulations were carried out on the Vienna Scientific Cluster (VSC). P.K. and G.M. acknowledge financial support for the experimental part from the German Research Foundation, SFB-TR6 project C4.
\newpage

\section{Supplementary Material}

In this Supplementary Information we provide some additional technical details about the experiment and the computation of diffusion constants and the identification of defects, as well as a more detailed comparison of our results with the predictions of elasticity theory.

\section{Experimental Details}

The monolayer consists of polystyrene beads with (4.5 $\mu$m diameter) doped with iron oxide (Dynabeads\textsuperscript{\textregistered} 4.5, Invitrogen) making them superparamagnetic. The beads are sedimented by gravity to an otherwise interaction free interface of a hanging droplet spanned by surface tension in a top sealed cylindrical hole (4 mm in diameter) of an optical cuvette. The thermal activation height (out of plane motion) is less than 20 nm and is therefore neglected. Computer controlled regulation loops counting particle numbers and measuring their size relative to focal plane guarantee the flatness of the water-air interface. This is done by adjusting the volume of the droplet in sub-nanoliter units with a micro-syringe driven by a micro-stage with a frequency of $~$0.1 Hz. To align the whole setup with respect to gravity the experiment is mounted on a flexible tripod steered by an inclination sensor. This way changes in inclination are suppressed below $10^{-6}$ rad and spatial density variations are less than $0.1\%$ in the field of view. The monitored area is $863 \times 645$ $\mu$m$^2$  recorded with a CCD-camera of  $1392 \times 1040$ pixels (Marlin F 145-B), containing about $3000$ particles, whereas the whole monolayer consists of up to $5\times10^4$ particles. An elaborated description of the experimental setup can be found in \cite{KEIM_RSI_2009}. 

\section{Numerical Details} 

The simulated system consists of $M = 52 \times 60 = 3120$ particles in an almost quadratic simulations box with periodic boundary conditions. A cutoff of $r_c = 5.0 a$ for the interactions was used together with cell lists to accelerate the simulations. The system is studied with Metropolis Monte Carlo simulations at constant particle number, box size and temperature. Trial moves consist of single particle displacements carried out with a maximum displacement size selected to obtain an average acceptance probability of about 50\%. The simulation time scale is mapped to the physical time scale by comparing the self diffusion constants in the liquid state obtained from simulations and experiments (see Supplementary Information). In all our experiments and simulations $\Gamma = 160$ leading to a Young's modulus of $K=1.258 \Gamma = 201.3$ \cite{YOUNGMODULUS}.

\section{Diffusion}

To monitor the long time diffusion of the defect strings, we determine the mean square displacement $\langle \Delta r(t)^2 \rangle = \langle [r(t)-r(0)]^2\rangle$ of the center of mass $r$ of the string endpoints. The diffusion constant $D_{\text{\rm string}}$ is estimated by least square fit to the relation $\langle \Delta r(t)^2 \rangle=4D_{\text{\rm string}}t$ giving the diffusion constants shown in Fig. 2 of the main text. To relate the physical time and the time scale $\tau$ of the simulation corresponding to one Monte Carlo sweep, we have determined the diffusion constant of a particle in the fluid, which is also known experimentally \cite{MARET1997,HYDRODYNAMIC}. For $M=780$ particles and $\Gamma = 40$, the Monte Carlo simulation yields a diffusion constant of $D = 1.69 \pm 0.19 \times 10^{-6}\,  a_0^2/\tau$.  From the experimental diffusion constant $D_{\textit{\rm exp}} = 0.11\, \mu{\rm  m}^2/{\rm s}$ one obtains $\tau=a_0^2/(0.11\, \mu{\rm m}^2/{\rm s})$ for the time scale of the simulation. 

\section{Defect identifications} 

A quantitative description of the dynamics of defect strings requires the accurate identification and location of dislocations and vacancies. Here, we identify dislocations as pairs of particles with $5$ and $7$ neighbors, respectively \cite{MARET}. Neighbor numbers are determined using a Voronoi analysis \cite{VORONOI}. The position of a dislocation is defined to be at the center of mass of the $5$- and $7$-coordinated particles and the Burgers vector of the dislocation is orthogonal to the vector pointing from the $5$- to the $7$-coordinated particle. Single vacancies can, in principle, also be identified from the position of the dislocations associated with them \cite{LDSoftMatter2}. However, if vacancies form a cluster, some dislocations annihilate and the individual positions of the defects can not be resolved any longer. Here, we identify vacancies based on a underlying reference lattice consisting of a  perfect triangular crystal with the density of the defect free crystal. The defect identification consist of the following steps: Each particle is first assigned to its closest perfect lattice site. The perfect lattice is then positioned to minimize the sum of distances between particles and the assigned lattices sites. With this step, the lattice follows the center-of-mass motion of the crystal. The positions of the vacancies are then identified as the position of unoccupied lattice sites. This procedure to locate vacancies is robust and the number of vacancies is constant throughout the simulation even if the vacancies assemble into a string.

As explained in the main text, the dynamics of vacancy strings consist of long gliding periods interrupted by rotation events at which the direction of motion of the string changes. The rotation rate $k_{\rm rot} = N_{\rm rot}/t$ is  the number $N_{\rm rot}$ of rotation events  that have occurred during the time $t$. To determine the rotation rate, we identify the orientation of the strings at each time step and detect rotation events based on changes in the string orientation. The string orientation is defined by the relative position of the dislocations that terminate the strings and it coincides with the direction of the Burgers vectors. The string orientation is undefined at the rotation points, where the string can change between two orientations. 

\section{Dislocation interaction}

According to linear elasticity theory, the interaction energy of two dislocations with Burgers vectors ${\bf b}_1$ and ${\bf b}_2$ and separated by ${\bf R}$ is given by \cite{NABARRO}
\begin{equation}
\beta F = -\frac{K}{4 \pi} \left[({\bf b}_1 \cdot {\bf b}_2) \ln R - \frac{({\bf b}_1 \cdot {\bf R})({\bf b}_2 \cdot {\bf R})}{R^2}\right],
\end{equation}
where $K$ is Young's modulus and $\beta = 1/k_{\rm B}T$. In a coordinate system with $x$-axis in the direction of motion, the separation vector for a string of $N$ vacancies can be written as ${\bf R} = (s,N \sqrt{3}/2)$ and the Burgers vectors are given ${\bf b}_1 = (-1,0)$, ${\bf b}_2 = (1,0)$. Using the dimensionless variable $x=2s/(\sqrt{3}N)$ one then obtains
\begin{equation}
\label{equ:endtoendscaled}
\beta F(x) = \frac{K}{8 \pi} \left(\frac{1-x^2}{1+x^2} +  \ln \frac{1+x^2}{2} \right),
\end{equation}
where the free energy has been shifted to vanish at its minimum value. This function has the shape of a symmetric double well with minima located at 
at $x = \pm 1$. The minima are separated by a barrier of height $h = (K/8\pi) (1-\ln 2)$ at $x=0$. As shown in Fig. 3 of the main paper, the effective dislocation interactions extracted from our experiments and simulations display the double-well form predicted by elasticity theory for sufficiently long strings. A more detailed comparison is provided in the following. 

\begin{figure}[t]
\centerline{\includegraphics[width=7cm]{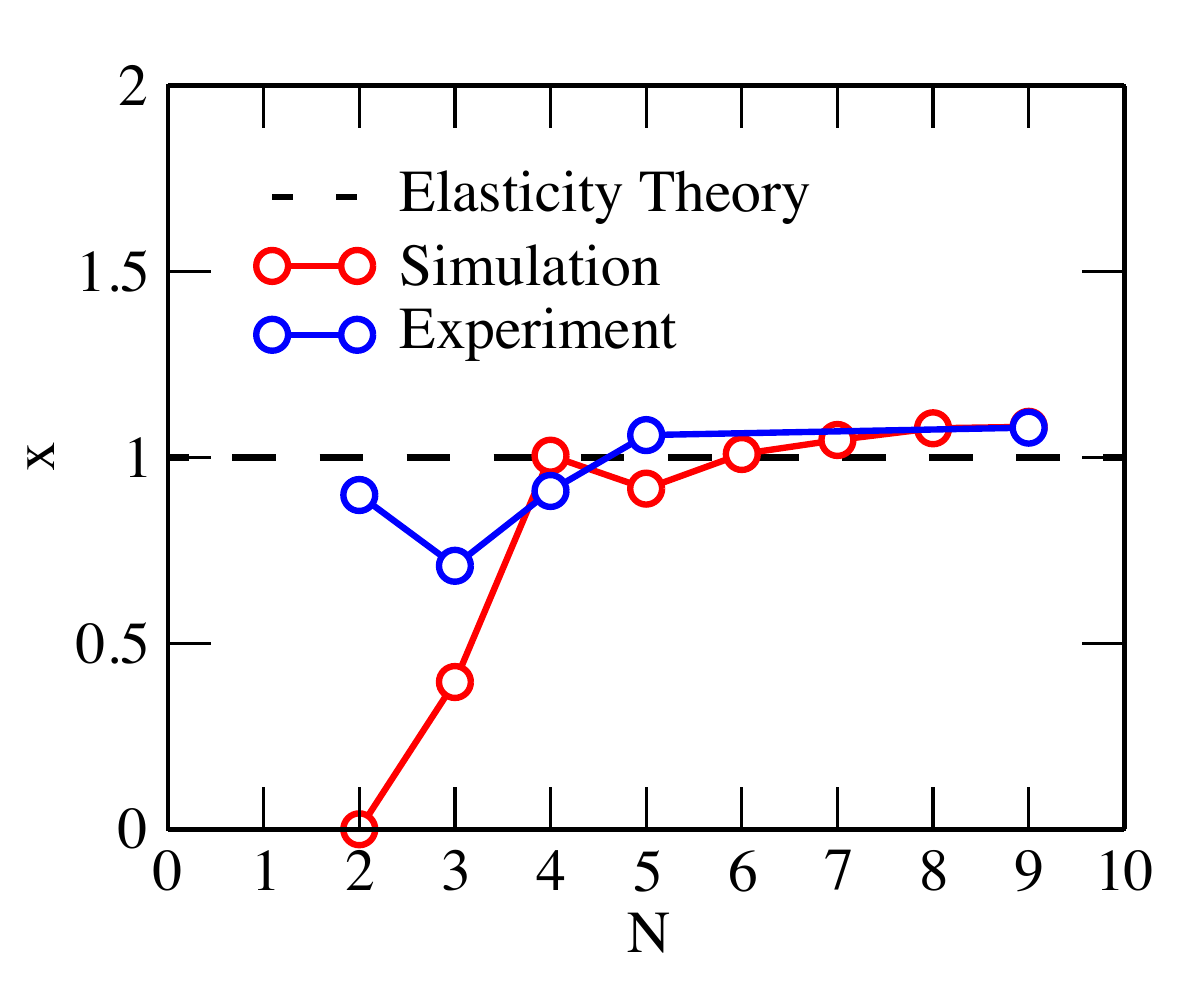}}
\caption{Position $x$ of the local minima as a function of string length $N$ obtained from simulations (red) and experiments (blue). For $N>3$ the prediction of elasticity theory (black horizontal line) is in excellent agreement with experiment and simulations.}
\label{fig:relativeerrors}
\end{figure}

The positions $x$ of the minima obtained from the experiment and simulations are depicted in Fig. \ref{fig:relativeerrors} together with the prediction of elasticity theory shown as horizontal line. For strings of more than $N>3$ vacancies, both experiment and simulation agree very well with elasticity theory. For short strings with $N \le 3$, however, the double-well form disappears and the minimum shifts to $x=0$.

\begin{figure}[t]
\centerline{\includegraphics[width=7cm]{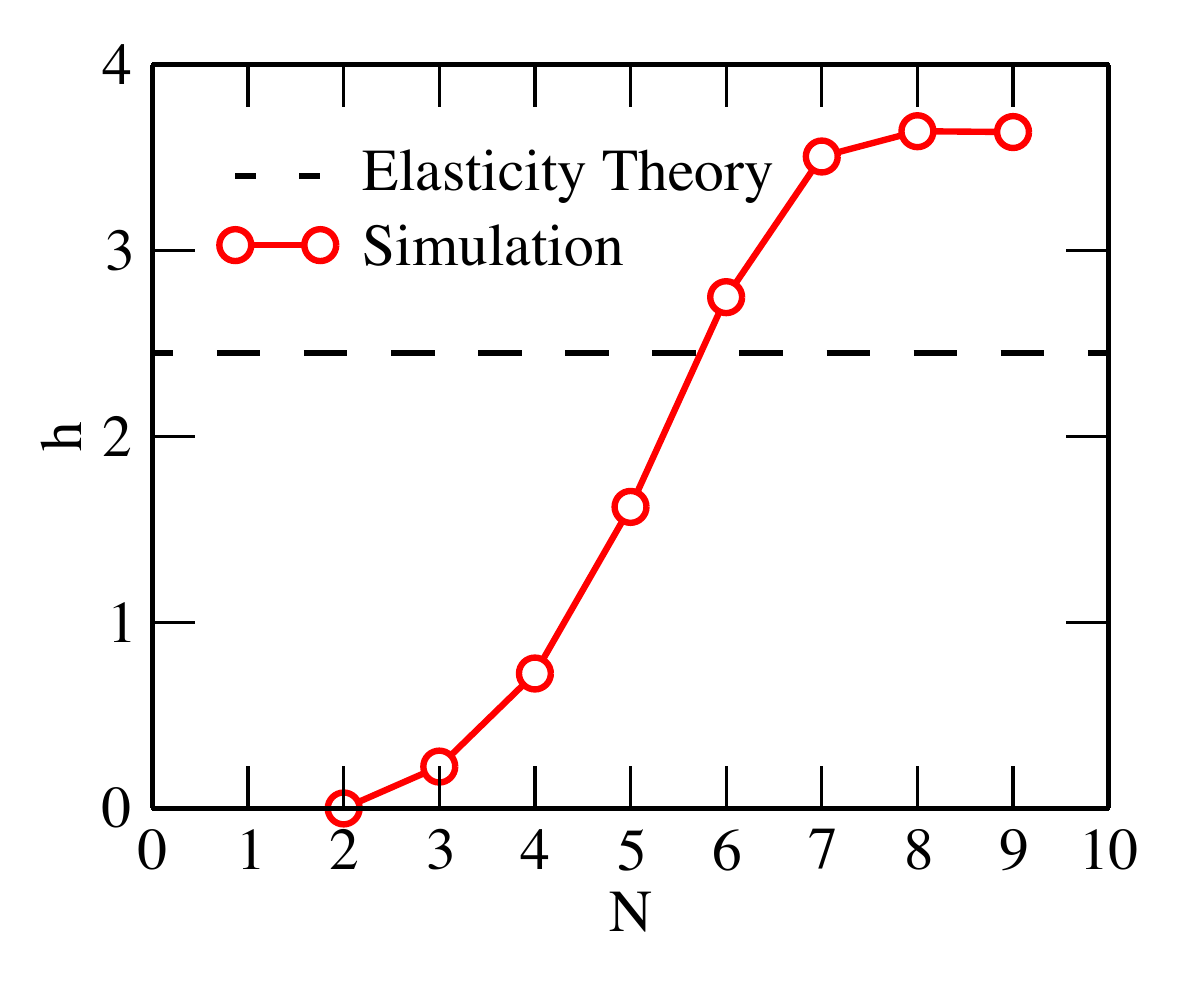}}
\caption{Barrier height $h$ obtained as a function of the string length $N$ from simulations (red) compared to the predictions of elasticity theory (black horizontal line). While for short strings the barrier disappears, for long strings it converges to a constant value which exceeds the prediction of elasticity theory by about 50 \%.}
\label{fig:relativeerrors_barrier}
\end{figure}

Barrier heights $h$ obtained for different string lengths $N$ are depicted in Fig. \ref{fig:relativeerrors_barrier}. Since the statistics of the experimental measurements are insufficient for an accurate determination of barrier heights, only simulation results are shown. While elasticity theory predicts a barrier height of $h=2.45 k_{\rm B} T$ independent of string length, the barrier height determined in our simulations vanishes for $N=2$ and then grows with string length until it converges to a constant value for strings consisting of more than about $N=7$ vacancies. The barrier height obtained for long strings exceeds the elasticity theory value by about 50\%. While one expects elasticity theory to break down at small distances, the origin of this discrepancy observed in the long string limit is unclear and might be due to non-linear interactions. 
\newpage

\bibliographystyle{prsty}

\end{document}